\documentclass[a4paper,12pt]{article}
\begin{document}
\thispagestyle{empty}
  \title{Interplay of Fulde-Ferrell-Larkin-Ovchinnikov and Vortex states 
in two-dimensional Superconductors}
\author{
U.Klein$^{1}$, D.Rainer$^{2}$, and H.Shimahara$^{3}$ \\
\small {$^{1}$Institut f\"ur Theoretische  Physik, 
       Universit\"at Linz, A-4040 Linz-Auhof, Austria}\\
\small {$^{2}$Physikalisches Institut, Universit\"at Bayreuth, D-95440 
       Bayreuth, Germany}\\
\small {$^{3}$Faculty of Science, Hiroshima University, 
       Higashi-Hiroshima 739-8526, Japan}
       }
\maketitle
\begin{abstract}
Clean superconductors with weakly coupled conducting planes have 
been suggested as promising candidates for observing the Fulde-Ferrell-
Larkin-Ovchinnikov (FFLO) state. We consider here a layered superconductor
in a magnetic field of arbitrary orientation with respect to the 
conducting plane. In this case there is competition of 
Pauli spin-pair-breaking effects, favoring the FFLO state, and 
orbital-pair-breaking effects, favouring the Abrikosov vortex
state. In previous work, phase transitions to phases with pairing in 
Landau levels with quantum numbers $n > 0$ have been predicted. Here, 
we calculate the actual structure of the stable 
states below $H_{c2}$ by minimizing the free energy. We 
find new order parameter structures differing from both the traditional 
Abrikosov and FFLO solutions. These include two-dimensional periodic 
structures with several zeros of the order parameter, as well as 
quasi-one-dimensional structures consisting of vortex chains separated 
by FFLO domains. We discuss the limit of high $n$, where some interesting 
but yet unsolved questions appear.
\end{abstract}
\begin{flushleft}
\hspace{2.0cm}PACS numbers: 74.25.Ha, 74.80.-g, 74.80.Dm 
\end{flushleft}
\pagebreak[4]

\section[intro]{Introduction}
\label{sec:intro}

Suppression of superconductivity by a magnetic field is a 
consequence of its interaction with either the magnetic moment due 
to the orbital angular momentum~\cite{GINZBURG} or due to the 
spin~\cite{CLOGSTON} of the electrons. The recent discovery 
of several classes of layered compounds with
nearly decoupled superconducting planes has renewed the interest in 
the interplay of spin and orbital pair breaking mechanisms. These 
include high-$T_{\rm c}$ cuprate superconductors, organic 
superconductors with very large upper critical fields, such as
$(\mbox{TMTSF})_2\mbox{PF}_6$~\cite{LNDC}, and hybrid ruthenate-cuprate
compounds like $\mbox{RuSr}_2\mbox{GdCu}_2\mbox{O}_8$~\cite{BTNBGBKNSA,WSP}.
The orbital pair breaking effect, described by a vector potential 
$\vec{A}$, is usually much larger than the spin effect and dominates most 
of the phenomena observed in magnetic fields, e.g. the structure of 
vortices in type II superconductors. Nevertheless, the spin effect has 
attracted considerable interest~\cite{CLOGSTON,CHANDRA}; 
it has soon be realized that its relative importance can be greatly 
enhanced by reducing the spatial extension of samples in a direction 
perpendicular to the external field. For the case of an 
almost two-dimensional superconductor with applied field parallel to 
the conducting planes, the orbital upper critical field is extremely 
high and the spin pair-breaking mechanism becomes the dominant one.

Then, if the orbital pair breaking mechanism can be completely neglected,
the \emph{homogeneous} superconducting state at $T=0$ becomes 
energetically  favourable~\cite{CLOGSTON,CHANDRA} at an 
upper critical field given by $\mu H_c=\Delta_0/\sqrt{2}$, where 
$\Delta_0$ is the BCS gap at $T=0$ (Chandrasekhar-Clogston 
limit). Later, a spatially \emph{inhomogeneous} state (FFLO 
state) with higher critical field was theoretically predicted 
by Fulde and Ferrell~\cite{FULFER}, and Larkin and 
Ovchinnikov~\cite{LAROVC}. In the FFLO state the order 
parameter performs one-dimensional spatial oscillations, which 
reduce the pair-breaking effect of the external field. 
Other theoretical studies on the FFLO state include the 
relation to the mixed state~\cite{GRUGUN1,TTGWLGSMPO}, the 
relevance of the shape of the Fermi 
surface~\cite{AODIFU,SHIMAFS}, heavy-fermion 
superconductivity~\cite{YINMAKI}, one-dimensional 
systems~\cite{DUPUIS}, the vicinity of the tricritical 
point~\cite{BUZKUL}, and a calculation of the lower critical 
field~\cite{BURRAI}; to mention only a few. Recently, 
it has been shown by one of us~\cite{SHIMAH}, that for 
superconductors with cylindrical (circular) Fermi 
surface one finds, at low temperature T, several different 
types of two-dimensional periodic 
structures with lower free energy than the traditional 
FFLO-state. 
Experimentally, while indications of spin paramagnetic 
effects have been observed, no clear evidence for an 
FFLO-like  phase has 
been obtained so far. The recent discovery of several classes 
of compounds with nearly isolated conducting layers has, 
however, opened new possibilities to detect the FFLO state. 

In most treatments, either the ``pure'' orbital effect or the ``pure'' 
spin effect has been investigated. Here, we study, following  
Bulaevskii~\cite{BULAEVSKII}, a situation where 
both effects must be considered. The relative importance of the spin effect 
is enhanced by means of the following geometrical arrangement: Consider 
a quasi-two-dimensional thin 
film with a large component $\vec{H}_\parallel$ of the applied magnetic 
field parallel to the film, and a small perpendicular 
component $\vec{H}_\perp$. The smallness of the perpendicular component 
makes the spin effect comparable  with the (much stronger) 
orbital effect.  
The upper critical field for such a situation has first been 
calculated by Bulaevskii~\cite{BULAEVSKII} and later studied 
in the vicinity of the tricritical point~\cite{BUZKUL}. Still 
later, Bulaevskii's  results  were rediscovered in a different 
theoretical framework, and generalized to arbitrary temperatures 
and d-wave superconductivity by~\cite{SHIRAI}. 
The essential result is a non-monotonic 
critical field curve $B_{c2}(T)$ which  
consists of several pieces belonging to different values of 
the Landau quantum number $n$. Such a behavior is in contrast 
to standard Ginzburg-Landau (GL) as well as microscopic theory,
where the pair-wave function is always found to be in the 
lowest Landau level $n=0$ at $H_{c2}$. Obviously, the gap 
at the Fermi surface opened by $\vec{H}_\parallel$  
forces  the single electrons in pairing states with  
higher Landau levels $n>0$. We note that formally similar 
upper critical field curves may also occur in a   
different physical context, at extremely high fields 
when Landau quantization of the single electron levels must 
be taken into account~\cite{GRUGUN,TERAXI,LEBYAM,NOAKDO}.
For the present situation, Landau quantization of the single 
electron levels can be neglected and the quasiclassical 
approximation holds. 

The present communication addresses 
the question of the actual structure of the spatially 
inhomogeneous 
state below some part of the upper critical field curve 
characterized by a Landau quantum number $n$. This state is of 
interest because it is a consequence of a competition 
between two different pair breaking mechanisms. It is 
also of practical interest because the portion of the 
$H,T$ plane where it occurs may be easier accessible 
experimentally than the one corresponding to the ``pure'' 
FFLO state. Note also, that a perfect orientation of the 
external field is difficult to achieve; as a consequence, 
a small perpendicular component is always present even if 
one tries to avoid it. 
\section[theory]{Calculation and theoretical background}
\label{sec:theory}
We use as a framework the 
quasiclassical Eilenberger equations~\cite{EILE1},
properly generalized to include the coupling between 
electron spins and magnetic field~\cite{RAIN1}. The notation 
used in the present work  is the same as in Ref.~\cite{KLEIN1}. 
We restrict ourselves to pure 
superconductors with an isotropic gap and a circular 
Fermi-surface. The quasiclassical equations consist of the transport 
equations for the Greens functions $f\mbox{,}f^{\it+}\mbox{,}g$ and the 
self-consistency equations for the order parameter $\psi$ and the 
vector potential $\vec{A}$. The transport equations are given by:
\begin{equation}
  \label{eq:quasif}
\left[ 2\omega_s+\hbar \vec{v}_F(\hat{k})\left( \vec{\nabla}_R-
\imath(2e/\hbar c) \vec{A}\right) \right] 
f(\vec{R},\hat{k},\omega_s) = 2\psi(\vec{R}) g(\vec{R},\hat{k},\omega_s)\mbox{,}
\end{equation}
\begin{equation}
  \label{eq:quasifplus}
\left[ 2\omega_s - \hbar \vec{v}_F(\hat{k}) 
\left(
\vec{\nabla}_R+\imath(2e/\hbar c) \vec{A}      
\right)
 \right] 
f^{\it+}(\vec{R},\hat{k},\omega_s) = 2\psi^\ast(\vec{R}) 
g(\vec{R},\hat{k},\omega_s)\mbox{,}
\end{equation}
\begin{equation}
  \label{eq:quasig}
g(\vec{R},\hat{k},\omega_s) = \left( 1-
f(\vec{R},\hat{k},\omega_s) 
f^{\it+}(\vec{R},\hat{k},\omega_s) \right)^{1/2} \mbox{.}
\end{equation}
Here, the Fermi velocity is denoted by $\vec{v}_F(\hat{k})$, 
the Zeeman term is contained in the quantity $\omega_s=
\omega_l+\imath \mu B$, where $\omega_l$ are 
Matsubara frequencies, $\mu$ is the magnetic moment of the electron and $B$ is the magnitude of the induction. The 
self-consistency equations are given by
\begin{equation}
  \label{eq:scop}
\left(2 \pi k_B T  \sum_{l=0}^{N_D} \frac{1}{\omega_l}+ 
\log\left(T/T_c\right) \right) 
\psi(\vec{R}) = \pi k_B T  \sum_{l=0}^{N_D} 
\int d^2 \hat{k}^\prime \, \left[ f(\vec{R},\hat{k}^\prime,\omega_s) +
f(\vec{R},\hat{k}^\prime,\omega_s^\ast ) \right] 
\mbox{,}
\end{equation}
\begin{displaymath}
\vec{\nabla}_R \times \left( 
\vec{B}(\vec{R}) + 4\pi \vec{M}(\vec{R}) \right)= 
\end{displaymath}
\begin{equation}
  \label{eq:scvecpot}
\frac{16\pi^2ek_BTN(E_F)}{c} \sum_{l=0}^{N_D} 
\int \frac{d^2 \hat{k}^\prime}{4\pi} \, \vec{v}_F(\hat{k}^\prime)
\Im g(\vec{R},\hat{k}^\prime,\omega_s ) \mbox{,} 
\end{equation}
\begin{equation}
  \label{eq:magn}
\vec{M}(\vec{R})=
2\mu^2N(E_F)\vec{B}(\vec{R})-4\pi k_BTN(E_F)\mu \sum_{l=0}^{N_D}
\int \frac{d^2 \hat{k}^\prime}{4\pi} \Im g \frac{\vec{B}}{B} 
\mbox{,} 
\end{equation}
The magnetization $\vec{M}$ comes from the spins;
the second term in Eq.~(\ref{eq:magn}) is due to the interaction 
between spins and supercurrent and may be neglected in the 
high $\kappa$-limit. These equations have to be supplemented 
by an expression for the Gibbs free energy $G$. A functional 
$G$ leading  to the above Euler-Lagrange equations may be 
constructed as a straightforward generalization of Eilenbergers 
original expression~\cite{EILE1} and will not be written 
down here.

As a first step, the differential operator in the 
transport equations has to be inverted using standard  
methods~\cite{HELFWERT}. This calculation 
reproduces the equation for $B_{c2}$ 
obtained previously~\cite{BULAEVSKII,SHIRAI}, as well as 
the eigenfunctions of the linearized gap equation, which 
are given by
\begin{equation}
 \label{EF}
\phi_n(k,\vec{r})=\frac{A}{\sqrt{n!}} 
\left( \frac{(-1)}{\sqrt{2}} \right)^n
\exp\left(\frac{\imath}{\hbar}kx\right)
\exp\left(
-\frac{\kappa_\perp}{2\hbar}\left(y-y_k\right)^2 
\right)
H_n\left(\sqrt{\frac{\kappa_\perp}{\hbar}} 
\left(y-y_k\right) \right)
\mbox{.} 
\end{equation}

Here, $x,y$ are the coordinates in the superconducting 
layer, $n=0,1,2,\ldots$ is a discrete (Landau level)
quantum number, $k$ is a continuous quantum number,
and $H_n$ is the Hermite polynomial of order $n$.
The remaining quantities in (\ref{EF}) are given by 
$\kappa_\perp= \frac{2\left|e\right|}{c}B \sin{\Theta}$
and $y_k=k/\kappa_\perp$, where $\Theta$ is the (small) 
angle between the external magnetic field and the 
superconducting plane. The value of $A$ may be chosen to 
normalize the complete, orthogonal set of 
eigenfunctions~(\ref{EF}). The same form of the normal  
state vector potential as in previous work~\cite{SHIRAI,
KLEIN2} has been chosen.

We want to study the structure of the (stable) inhomogeneous
state below $B_{c2}$ for arbitrary $n$. For given 
temperature $T$, angle $\Theta$, and ratio $T_c/E_F$, the 
correct value of $n$ is obtained by solving numerically 
the equation for $B_{c2}$~\cite{SHIRAI}. Given this 
$n$, the order parameter below the corresponding branch 
of the $B_{c2}$ curve can be constructed as a linear 
combination 
of eigenfunctions belonging to the infinitely degenerate 
set~(\ref{EF}), which are labeled by the continuous quantum 
number $k$. We construct a general state with an order 
parameter which is quasi-periodic (periodic up to a phase 
factor) in  two dimensions. Abrikosov's method~\cite{ABRI}, 
used originally for $n=0$ and $T \sim T_c$, 
may be applied in a straightforward manner to the present case 
of arbitrary $n$. It leads  to an order parameter $\psi_n$ 
which is given by
\begin{eqnarray}
 \label{OP}
\psi_n(X,Y)&=&AC_n\sum_{m=-\infty}^{m=+\infty}\exp\left(-2 \pi \imath \frac{b}{a} \cos\alpha \frac{m(m+1)}{2} \right) \times
\nonumber\\
& &\cdot \exp\left(\frac{\imath}{\hbar}k_m \left(X+Y\cos\alpha  \right)  \right) h\left(\sin\alpha \left(Y-mb \right) \right)
\mbox{,} 
\end{eqnarray} 
where
\begin{equation}
 \label{HFORM}
h(x)= \frac{A}{\sqrt{n!}} 
\left( \frac{-1}{\sqrt{2}} \right)^n
\exp\left(-\frac{\kappa_\perp}{2\hbar}x^2 \right)
H_n\left(\sqrt{\frac{\kappa_\perp}{\hbar}} 
x \right) \mbox{.} 
\end{equation}
Here, $\alpha$ is the angle between the two primitive 
lattice vectors $\vec{a},\vec{b}$ (of length $a,b$) 
spanning the unit cell of the periodic structure we 
are interested in. An oblique system of coordinates 
$X,Y$ has been used in~(\ref{OP}) whose relation to  
Cartesian coordinates $x,y$ is given by 
$x=X+Y\cos\alpha \mbox{, } y=Y \sin\alpha$. In the course 
of the calculation leading to~(\ref{OP}), the continuous
numbers $k$ have been restricted to a discrete set
$k_m=(2 \pi \hbar /a)m$ with integer $m$. The flux due
to the perpendicular component $B_\perp=B\sin\Theta$ must be 
quantized; the corresponding fluxoid quantization 
condition has been used in the derivation of~(\ref{OP}). 
It takes the form $\kappa_\perp a b \sin\alpha=
2 \pi \hbar$ if we assume that each unit cell carries \emph{one} 
flux quantum. Periodic order parameters for arbitrary $n$ have been
reported previously~\cite{AKDOGI,RIESCHSCH} in the literature.

Among all possible structures~(\ref{OP}) the stable one, 
i.e. the one with the lowest free energy, must be found. 
To perform this task the quasiclassical equations have to be 
solved near $B_{c2}$ and the free energy has to be calculated up
to terms of fourth order in the order parameter magnitude. In contrast
to the GL region, the coherence lenght is finite at low $T$; 
therefore all powers of the order parameter gradient must be taken 
into account. This large   calculation, which generalizes   
Abrikosov's solution to arbitrary $T$, has first been performed by 
Eilenberger~\cite{EILE2} for the ordinary vortex lattice, i.e. for $n=0$. 
Later, Eilenbergers mixed state calculations have been generalized by 
Rammer and Pesch~\cite{RAMPESCH} to include strong coupling effects. The 
present calculation generalizes Eilenbergers 
work to arbitrary Landau quantum numbers $n$. It 
makes extensive use of calculational 
techniques~\cite{EILE3,HELFWERT,EILE2,DELRIEU} 
developed by previous workers in this field and comprises the 
following main steps: (1) Solution of  
the linearized transport equations, 
(2) expansion  of transport equations and free
energy for small order parameter, (3) solving Maxwells equation,
and (4) performing a large number of spatial and momentum 
integrations.
While all second order free energy contributions can be calculated 
exactly, one has to perform  an asymptotic approximation in integrals 
appearing in fourth order terms in order to obtain a simple final 
result. This approximation is similar to the one suggested by 
Delrieu~\cite{DELRIEU} and holds for not too low $T$ 
(see the discussion following Eq.(6) of Delrieu's 
paper~\cite{DELRIEU}). 

After a long  calculation, which will be described in more 
detail elsewhere, one obtains the following expansion (note 
that dimensionless units~\cite{KLEIN2} will be used in what 
follows) for the free energy $G$: 
\begin{equation}
  \label{eq:gexpansion}
  G=\bar{G}+\tilde{\alpha}^{2}\bar{G}^{(2)}+
\tilde{\alpha}^{4}\bar{G}^{(4)}
\end{equation}
Here, $\tilde{\alpha}^{2}$ denotes the spatial average of $\psi^2$, 
our small expansion parameter. The term $\bar{G}$ stems from 
the spatially constant part of the magnetic 
induction~\cite{EILE2}; it is unimportant in the present 
context since it does not depend on the structure variables 
$\alpha,a,b$. The second order contribution is given by 
\begin{equation}
  \label{eq:g2quer}
  \bar{G}^{(2)}=\ln t + t\int_0^\infty ds \frac{1}{\sinh(st)}
\left[1-\cos(\mu \bar{B}s )f_1(s^2\frac{\bar{B}_\perp}{2})  
\right] \mbox{,}  
\end{equation}
where $t=T/T_c$, and $\bar{B}$, $\bar{B}_\perp$ are spatially averaged
values of $B(\vec{R})$ and $B_\perp(\vec{R})$ respectively. The function
$f_1$ depends on the Landau quantum number $n$ and is given by 
\begin{equation}
  \label{eq:f1function}
  f_1(x)=\exp(-\frac{x}{2})L_n(x)
\mbox{,}
\end{equation}
where $L_n$ is the Laguerre polynomial of order $n$. The second order 
term (\ref{eq:g2quer}) is also independent of $\alpha,a,b$, but 
the equation $\bar{G}^{(2)}=0$ 
has to be solved in order to find the upper critical field 
and the corresponding value of $n$ (this equation for $B_{c2}$ agrees 
with the one given in~\cite{SHIRAI}).

The new information is contained in the fourth order term $\bar{G}^{(4)}$.
The magnetic part, say $\bar{G}^{(4)}_M$, of $\bar{G}^{(4)}$ is of 
order $1/\kappa^2$. This magnetic contribution collects all the terms 
related to Maxwells equation. If 
Eilenbergers, or Abrikosovs results are to be rederived, as a special case 
of the present theory for $\vec{H}_\parallel=0$ ($n=0$), \emph{all} 
contributions to the free energy, 
including $\bar{G}^{(4)}_M$, have, of course, to be taken into account (one 
can successfully check Eq.(\ref{eq:gexpansion}) against these classical 
results; the same is true for the FFLO upper critical field which may be 
derived by solving $\bar{G}^{(2)}=0$ in the limit $\vec{B}_\perp =0$ ). In the 
present communication we consider the high-$\kappa$ limit and neglect 
the contribution $\bar{G}^{(4)}_M$. This approximation 
is used by most workers on the FFLO state, in particular in 
Larkin and Ovchinnikov's original paper~\cite{LAROVC}. 
A complete treatment including this magnetic contribution 
as well as magnetization curves for the present geometry, 
will be reported in a separate publication~\cite{KLEIN4}. 
Then, the final result for the fourth order term is given by
\begin{equation}
  \label{eq:g4querf}
\bar{G}^{(4)}=\frac{t}{4} \sum_{l=-\infty}^{+\infty}
\sum_{j=-\infty}^{+\infty} f_1^2(x_{l,j}) S_{l,j}
\mbox{,} 
\end{equation}
where $f_1$ has been  defined in Eq.(\ref{eq:f1function}). 
The Matsubara sum $S_{l,j}$ is defined by
\begin{equation}
  \label{eq:S1matsum}
S_{l,j}= \sum_{i=0}^{N_D} 
\frac
{2 \omega_l^2+\bar{B}_\perp x_{l,j}/2}
{
\omega_l^2\left(\omega_l^2+\bar{B}_\perp x_{l,j}/2 \right)^{3/2}
}\mbox{,}  
\end{equation}
and $x_{l,j}$ is given by
\begin{equation}
  \label{eq:xlj}
x_{l,j}=\frac{\pi^2}{\sin^2\alpha}
\left(\frac{L}{a}\right)^2 l^2+
\left(\frac{a}{L}\right)^2 j^2-
2 \pi lj\frac{\cos\alpha}{\sin\alpha}
\mbox{.}
\end{equation}
The minimization with respect to the flux-line structure 
is done at \emph{fixed} $\bar{B}_\perp,\bar{B}_\parallel$ 
 (the minimization with respect to $\bar{B}_\perp,
\bar{B}_\parallel,\tilde{\alpha}$ has to be done in separate steps). 
Therefore, stability conditions should not be applied with 
respect to the variables $a/b$,~$\alpha$ but with respect to  
$a/L$,~$\alpha$, where $L$ is a magnetic length defined by 
$L^2=2/\bar{B}_\perp$.
Then, using the results for $a/L,~\alpha$, the ratio $a/b$ 
is finally determined with the help of the fluxoid quantization 
condition.

The fourth order term~(\ref{eq:g4querf}) has a simple 
interpretation. As is well known, near $T_c$ the conventional 
flux-line structure is completely determined by a single 
quantity, Abrikosov's geometrical factor 
$\beta_A=\bar{\psi^4}/\bar{\psi^2}^2$ (the bar denotes a spatial
average). If this quantity is computed using a general  
order parameter $\psi_n$, as given by Eq.~(\ref{OP}), 
one obtains, after a lengthy calculation
\begin{equation}
  \label{eq:betaa}
  \beta_{A,n}=\sum_{l=-\infty}^{+\infty}
\sum_{j=-\infty}^{+\infty} f_1^2(x_{l,j})
\mbox{.} 
\end{equation}
One notices that the free energy~(\ref{eq:g4querf})
differs from $\beta_{A,n}$ only by the temperature-dependent 
factor $S_{l,j}$ modifying each term in the double sum 
(for $n=0$, $\beta_{A,n}$ takes its lowest value $1.1596$ 
for a triangular lattice). The factor 
$\bar{B}_\perp x_{l,j}/2$ in $S_{l,j}$, (see 
Eq.~(\ref{eq:S1matsum}), vanishes in the GL-limit; it 
reflects nonlocal correlations in Eilenbergers  
transport equations and implies an additional ``microscopic''  
temperature dependence of the stable magnetic flux structure.  
One expects, that for higher $T$, the GL approximation
$\beta_A$ may be used as a first approximation for  
the complete free energy~(\ref{eq:g4querf}). It should also 
be mentioned that the electrons magnetic moment $\mu$, which 
is the typical spin pair breaking parameter, does not 
explicitly occur in Eq.~(\ref{eq:g4querf}) (but only in 
$\bar{G}$,~$\bar{G}^{(2)}$, and $\bar{G}^{(4)}_M$). The spin
effect influences, however, the free energy - and the resulting
 magnetic flux structure - in a very decisive way since 
it determines the value of the Landau quantum number $n$ in 
Eq.~(\ref{eq:g4querf}).
\section[results]{Results and discussion}
\label{sec:results}
To calculate numerically the stable flux structure the following
two steps have to be performed: (1) The equation $\bar{G}^{(2)}=0$ 
has to be solved for a given temperature $T$ and angle $\Theta$ to obtain   
the upper critical field and the corresponding value of $n$. (2) Using
these  values the stable flux structure has to be determined by 
calculating the absolute minimum of~(\ref{eq:g4querf}) with respect 
to $a/L$,~$\alpha$. Throughout the numerical calculations a value 
of $\mu=0.1$ ($\mu=(\pi/2) k_BT_c/E_f$ in the present system of units) 
has been used. The calculations have been performed (for various
$n$ in the range between $0$ and $30$) at two different 
temperatures $t=T/T_c=0.5$ and $t=0.2$, in order to investigate 
the influence of the above discussed low-temperature 
correlations.

The minimization procedure turned out to be more 
involved~\cite{PREUSSER} than one 
would expect from the simple form  of Eq.~(\ref{eq:g4querf}). 
One finds a large number of minima belonging either to 
different lattice structures or to different basis vectors of 
the same lattice. The number of minima 
increases rapidly with increasing Landau quantum number $n$. The quantity 
shown in all plots is actually $\bar{G}^{(4)}_N$, which is $\bar{G}^{(4)}$ 
multiplied by a factor of $16t^3/7\chi(3) \approx 2t^3$ (in order to 
normalize it for $t \rightarrow 1$ to the GL result). Fig.~1 shows 
a contour plot for $n=10,t=0.5$ of $\bar{G}^{(4)}_N$ in the most important part 
of  the $\alpha,a/L$-plane; higher values of $a/L$ need not be considered,
since they correspond to lattices with exchanged axes $a,b$. The minima 
of the free energy are visible as dark regions. Two different types of 
minima may be distinguished in Fig.1 (and similar plots). A minimum of 
the first type is represented by an isolated dark region and corresponds 
to an ``ordinary'' two-dimensional lattice (several equivalent points 
in the $\alpha,a/L$-plane, which have the same 
free energy and belong to the same lattice, but to different basis vectors, 
may be found in Fig.1). The second type of minimum can be found in the 
straight ($\alpha$-independent) dark regions in the lower part of Fig 1. 
Near these ``flat valleys'' the free energy depends 
mainly on $a/L$, a small dependence on $\alpha$ along some lines of 
fixed $a/L$ exists, but is invisible in Fig 1. The minima in these flat 
valleys will be referred to as ``quasi-one-dimensional'' lattices, because 
of the small energy barrier along the lines of constant $a/L$; a more
detailed discussion will be given below.

The phases with $n=0,1,2$ occupy a relatively large portion of the phase 
diagram~\cite{BULAEVSKII,SHIRAI} and are therefore of particular interest 
in view of a possible experimental detection of these new states. For the 
lowest Landau index $n=0$, we recovered at both temperatures, as expected, 
the conventional triangular vortex lattice as stable state (global minimum of
the free energy). To create a state with $n=1$, a possible 
choice for the tilting angle of the external magnetic field is  
$\Theta=0.3$ (degree) at $t=0.5$; another possible choice is 
$\Theta=1.2$ at $t=0.2$. The stable structure found for $n=1$ 
is, at both temperatures, of the quasi-one-dimensional type. A contour
plot of the magnitude of the order parameter is shown in Fig.~2; the 
structure  consists of chains of vortices separated by a single line 
of vanishing order parameter. The latter may be considered as a fragment 
of the one-dimensional FFLO-state; for higher $n$ local minima 
appear whose  order parameters look similar to Fig.2 except that $n$ lines 
of order parameter zeros appear between the vortex chains. 

Let us discuss these quasi-one-dimensional structures - which have been 
verified as local, but not necessarily global, minima for many higher 
Landau indices in the range $n<30$ - in more detail. 
They consist of two subsystems, vortices and one-dimensional 
FFLO-type oscillations. Both of these are of the ``ordinary'' type, i.e. 
they are basically unaffected by their mutual interaction; the phase change 
when encircling a vortex is $2\pi$ and the wave length as well as the 
behavior of the phase of the one-dimensional periodic structure agree with 
that of the one-dimensional FFLO state. For all of these states, 
the coupling energy between neighboring vortex chains is small. 
As a consequence, the energy barrier 
preventing motion of the vortex chains in the direction parallel to the
chains is very small (for $n=1$ the $\alpha$-dependence shows up in the 
sixth digit of the free energy). Thus, even small fluctuations 
will destroy the periodicity in 
the direction parallel to the chains, making the material effectively 
one-dimensional. These decoupling of the vortex chains leads to a  
quasi-continuum of unit-cells (similar to the one shown in the lower 
part of Fig.~1 for $n=10$) which is located near $a/L=1.03$ for $n=1$. 
Experimentally, one expects a pronounced change in transport 
properties if, by decreasing the angle $\theta$, the system makes 
a transition from $n=0$ to $n=1$.  

As regards the next higher Landau level index, $n=2$, stability calculations 
have been performed for the two sets of parameters $\Theta=0.17,\,t=0.5$ and
$\Theta=0.7,\,t=0.2$. At $t=0.5$ a triangular lattice is realized as 
the state of lowest free energy. At $t=0.2$ a slightly distorted triangular 
lattice ($\alpha=60.05$) is realized. The distortions are due to the 
microscopic factor $S_{l,j}$ in Eq.~(\ref{eq:g4querf}). Thus, the influence 
of the microscopic correlations on these low-$n$ states is visible but very 
small. The order parameter for $n=2$ is shown in Fig.~3. 
The triangular unit cell contains three order parameter zeros. These zeros 
have different vorticity; the phase change when encircling a zero is 
$+2\pi$ for two of the vortices and $-2\pi$ for the third. Recall 
that, for arbitrary $n$, the unit cell as a whole carries exactly one 
flux quantum.

For $n=3$ at $t=0.5$ we found again a quasi-one-dimensional stable state. 
The order parameter for $n=3$ looks similar to Fig.~2, except for  
\emph{three} dark lines appearing  between the vortex chains. No stable 
quasi-one-dimensional states have been found for higher $n$ ($n>3$). In the 
interval $3<n<10$ the global free energy minima are given by various 
types of two-dimensional lattices, including triangular, quadratic and oblique 
structures. With increasing $n$ the influence of the microscopic correlations 
on the lattice structure increases; the stable state does not only depend 
on $n$ but may change between $t=0.5$ and $t=0.2$ (keeping $n$ constant).

The phases with higher $n$ correspond to very small values of $\Theta$ (e.g.
for $n=8$ and $t=0.5$ one has $\Theta=0.05$). One expects the 
resulting stable states 
to be already very similar to the FFLO state, which corresponds to the limit 
$n \rightarrow \infty$, $\Theta=0$ ($\bar{B}_\perp=0$). This FFLO-limit 
$n\rightarrow\infty$ of the present theory is a subtle step. An analytical 
limit of the calculation cannot be performed since in
the present formulation extensive use has been made of the fluxoid 
quantization condition, which becomes meaningless if no 
perpendicular component exists. Nevertheless, one expects for 
general reasons of continuity, that the present theory comprises 
the FFLO theory in the sense, that, with increasing $n$, the 
numerical results become more and 
more similar to the FFLO state. Thus, for $t=0.5$ we expected, as a  
natural candidate for a smooth transition to the FFLO state, the 
quasi-one-dimensional state to be stable for large $n$. At $t=0.2$ we 
expected, in view of~\cite{SHIMAH}, a two-dimensional structure to be stable
for large $n$.

However, contrary to our expectation, no single stable lattice could 
be found in our calculations which have been performed up to $n=30$ 
for $t=0.5$ and $n=20$ for $t=0.2$. Contour plots of the free energy look 
similar to Fig.~1 but with an increased number of local minima at 
higher $n$. At both temperatures the transition from n to n+1 (keeping 
$t$ fixed and varying $\Theta$) leads generally not to the same but to 
a \emph{different} lattice; among these stable states we found all kinds 
of oblique, triangular and rectangular two-dimensional structures. 
Quasi-one dimensional states, which are stable at n=1,3, have also been 
found as local minima in these high-n calculations, but had always higher 
energy than the minima corresponding to two-dimensional states. 
The difference in free energy of the different local minima decreases 
with increasing n. 

At present, we are unable to decide whether this strange vortex-liquid 
like limiting behavior, which would imply the instability of the FFLO 
state, is an artifact of our calculations or corresponds to 
reality. There are some arguments in favor of the latter possibility: 
Adding a perpendicular component of the magnetic field, no matter how 
small, implies flux quantization; this changes  the conditions which 
determine the stable state in a discontinuous way. Further, the size  of 
the unit cell and the number of vortices (with different  vorticity) per 
unit cell increases with $n$. It seems reasonable to assume that the  
energy differences between  different vortex arrangements (lattice 
structures) become small for large $n$. More work is required to settle 
this question. 
\section[conclusion]{Conclusion}
\label{sec:conclusion}
We considered a geometrical arrangement 
with competition of orbital and spin pair breaking effects
in a two-dimensional superconducting state. This arrangement 
comprises the traditional vortex state (for $B_\parallel=0$) 
as well as the FFLO state (for $B_\perp=0$). Minimizing 
the free energy we found 
several new structures below $B_{c2}$ with pairing states in 
Landau levels $n>0$. These include two-dimensional
structures with vortices of different  vorticity as well
as mixtures of one-dimensional periodic order parameter 
oscillations and vortex chains. The latter show a fluid-like behavior, 
as regards vortex motion in the direction parallel to the chains. 
This  feature could be used, besides more local spectral 
techniques, to identify this new state experimentally. Topics 
to be treated in future work include the high-$n$ limit, contributions 
from finite values of $\kappa$, magnetization curves, a more 
systematic account of unit cell structures, and the transitions 
between pairing states belonging to different  $n$.

\pagebreak 
\newpage
\begin{center}
{\bf FIGURE CAPTIONS}  
\end{center}
\vspace{1cm}
\begin{enumerate}
\item Contour plot of the fourth order free energy 
contribution $\bar{G}^{(4)}_N$ for $n=10$,~$t=0.5$ as a 
function of $\alpha,a/L$. Minima of free energy must be sought in 
dark regions.        
\item Contour plot of $\left| \psi \right|^2$ as a function of
$x,y$. Stable structure for Landau level $n=1$ at $t=0.5$
with lattice parameters $a/b=2.75$,~$\alpha=69.83$. A small energy 
barrier exists against  translation of the vortex chains relative to 
each other.
\item Contour plot of $\left| \psi \right|^2$ as a function of
$x,y$. Stable structure for Landau level $n=2$ at $t=0.5$
with lattice parameters $a/b=1.00$,~$\alpha=60.00$
~($a/L=1.90$). The unit cell carries one flux quantum and contains 
three order parameter zeros.
\end{enumerate}

\begin{thebibliography}{99}
%
\bibitem{GINZBURG} V.L. Ginzburg, {\it Sov.Phys.JETP} {\bf 4} 153 (1957)
%
\bibitem{CLOGSTON} A.M. Clogston, {\it Phys. Rev. Lett.} {\bf 9} 266 (1962)
%
\bibitem{LNDC} I.J. Lee, M.J. Naughton, G.M. Danner and P.M. Chaikin, 
{\it Phys. Rev. Lett.} {\bf 78} 3555 (1997)
%
\bibitem{BTNBGBKNSA} C. Bernhard et al., preprint cond-mat/9901084, 1999
%
\bibitem{WSP} R. Weht, A.B. Shick, and W.E. Pickett, preprint 
cond-mat/9903210, 1999
%
\bibitem{CHANDRA} B.S. Chandrasekhar, {\it Appl. Phys. Lett.} {\bf 1} 7 (1962)
%
\bibitem{FULFER} P.Fulde and R.A. Ferrell, {\it Phys. Rev.} {\bf 135} A550 (1964)
%
\bibitem{LAROVC} A.I. Larkin and Y.N. Ovchinnikov, {\it Sov. Phys. JETP} {\bf 28} 1200 (1969)
%
\bibitem{GRUGUN1} L.W.Gruenberg and L.Gunther, {\it Phys. Rev. Lett} 
{\bf 16} 996 (1966)
%
\bibitem{TTGWLGSMPO} M.Tachiki, S.Takahashi, P.Gegenwart, M.Weiden, 
M.Lang, C.Geibel, F.Steglich, R.Modler, C.Paulsen, Y.Onuki,
{\it Z. Physik} {\bf B100} 369 (1996) 
%
\bibitem{AODIFU} K.Aoi, W.Dieterich, and P.Fulde, {\it Z. Physik} {\bf 267} 
223 (1973)
%
\bibitem{SHIMAFS} H. Shimahara, {\it J. Phys. Soc. Jpn.} 
{\bf 66} 541 (1997)
%
\bibitem{YINMAKI} G.Yin and K.Maki, {\it Phys. Rev. }{\bf B48} 650 (1993)
%
\bibitem{DUPUIS} N.Dupuis, {\it Phys. Rev. }{\bf B51} 9074 (1995)
%
\bibitem{BUZKUL} A.I.Buzdin and M.L.Kulic, {\it J.\,Low\,Temp.\,Phys.} {\bf 54} 203 (1984)
%
\bibitem{BURRAI} H.Burkhardt and D.Rainer, {\it Ann. Physik} {\bf 3} 181 (1994)
%
\bibitem{SHIMAH} H.Shimahara, {\it J. Phys. Soc. Jpn.} 
{\bf 67} 736 (1998)
%
\bibitem{BULAEVSKII} L.N. Bulaevskii, {\it Sov. Phys. JETP} {\bf 38} 634 (1974)
%
\bibitem{SHIRAI} H.Shimahara and D.Rainer, {\it J. Phys. Soc. Jpn} {\bf 66} 
3591 (1997)
%
\bibitem{GRUGUN} L.W.Gruenberg and L.Gunther, {\it Phys. Rev} {\bf 176} 
606 (1968)
%
\bibitem{TERAXI} Z.Tesanovic, M.Rasolt, and L.Xing, {\it Phys. Rev. Lett.}
{\bf 63} 2425 (1989)
%
\bibitem{LEBYAM} A.G.Lebed, K.Yamaji, {\it Phys. Rev. Lett} 
{\bf 80} 2697 (1998)
%
\bibitem{NOAKDO} M.R.Norman, H.Akera, and A.H.MacDonald, 
{\it Physica} {\bf C196} 43 (1992)
%
\bibitem{EILE1} G.Eilenberger, {\it Z. Physik} {\bf 214} 195 (1968)
%
\bibitem{RAIN1} J.A.X. Alexander, T.P. Orlando, D. Rainer, and
P.M. Tedrow, {\it Phys. Rev. }{\bf B31} 5811 (1985)
%
\bibitem{KLEIN1} U. Klein, {\it Phys. Rev. } {\bf B40} 6601 
(1989)
%
\bibitem{HELFWERT} E.Helfand and N.R. Werthamer, {\it Phys. Rev. }{\bf 147} 
288 (1966)
%
\bibitem{KLEIN2} U. Klein, {\it J. Low Temp. Phys. } {\bf 69} 1 (1987)
%
\bibitem{ABRI} A.A.Abrikosov, {\it Sov.Phys.JETP} {\bf 5} 1174 (1957)
%
\bibitem{AKDOGI} H. Akera, A.H. MacDonald, and S.M. Girvin 
{\it Phys. Rev. Lett. }{\bf 67} 2375 (1991)
%
\bibitem{RIESCHSCH} C.T. Rieck, K. Scharnberg, and N. Schopohl 
{\it J. Low Temp. Phys. }{\bf 84} 381 (1991)
%
\bibitem{EILE2} G. Eilenberger, {\it Phys. Rev. } {\bf 153} 
584 (1967)
%
\bibitem{RAMPESCH} J. Rammer and W.Pesch {\it J. Low Temp. Phys.}
{\bf 77} 235 (1989)
%
\bibitem{EILE3} G. Eilenberger, {\it Z. Physik }
{\bf 180} 32 (1964)
%
\bibitem{DELRIEU} J.M. Delrieu, {\it J.\,Low\,Temp.\,Phys.} 
{\bf 6} 197 (1972)
%
\bibitem{KLEIN4} U.Klein, to be published
%
\bibitem{PREUSSER} The free Unix program ``xfarbe'' by A.Preusser 
has been used to perform the contour plots and calculate the 
minima.
%
\bibitem{TERAXI2} Z.Tesanovic, M.Rasolt, and L.Xing, {\it Phys. Rev.}{\bf B43} 288 (1991)
%

\end{thebibliography}
\end{document}